\documentclass[12pt]{article}
\usepackage{graphicx}
\begin{document}

{\bf The Penna Model of Biological Aging}

\bigskip

D. Stauffer

\bigskip
Institute for Theoretical Physics, Cologne University, 

D-50923 K\"oln, Euroland

\bigskip

Abstract: This review deals with computer simulation of biological ageing,
particularly with the Penna model of 1995.

\bigskip

\section{Introduction}

``There is no shortage of theories of ageing'' (Partridge 2007). Indeed it 
has been claimed that there as many aging theories as there are aging theorists.
This claim is inaccurate since this author alone simulated three different 
types of aging theories.

However, for this journal we are interested in computer simulations of 
biological aging,  and then the number of candidates goes down drastically. 
I do not claim that only those theories which have been simulated extensively
are correct, and that the other theories are wrong; I merely report on what
readers of this journal presumably are interested in. Thus in the following 
chapters I review recent progress on the most widespread model of biological
ageing, which was invented by Thadeu Penna before he was 30 years old
(Penna 1995). At the end, other models will be reviewed more briefly. 
There are many review articles and two books (Moss de Oliveira et al 1999,
Stauffer et al 2006) on this subject; thus we summarise here the research
with emphasis on recent developments.

\section{Reality}

Every year we age by one year if we do not die, and in middle age the mortality
increases roughly exponentially with age $a$, 
$$ {\rm mortality} \propto e^{ba} \quad , \eqno (1a)$$
as observed by Gompertz in 1825. This mortality can be defined as
$$ q = [S(a)-S(a+1)]/S(a) \eqno (1b) $$
where $a$ is the age (usually measured in years for humans) and $S(a)$ the 
number of people surviving from birth to age $a$;  this $q$ is the fraction of 
people who reach age $a$ and die before they reach age $a+1$. The Gompertz law
describes reality better if instead of this $q$ one looks at the mortality
function (or force of mortality) 
$$ \mu = - d \ln(S(a))/da  \eqno(1c)$$
since this age derivative is not based on one year as the only relevant time
interval and can become larger than one. If only life-tables for $S(a)$ based
on years are available, one may approximate this mortality function as
$$ \mu \simeq \ln[S(a)/S(a+1)] \quad .\eqno (1d)$$ 
Both expressions for $\mu$ mathematically can go to infinity for old age, as 
required by the Gompertz law, while $q$ by definition cannot exceed unity.

Many people want to become young again, and when for some flies a mortality
plateau was seen at very old age, claims were made that also for humans 
the mortality has a maximum somewhat below 100 years. However, the facts
of Fig.1 are less optimistic, and only above 110 years human mortality 
might reach a constant (Robine and Vaupel 2001). For the oldest Swedish men
the mortality function $\mu$ in Fig.1 is above 1, invalidating the earlier
and then very good fit (Thatcher et al 1998)
$$ \mu = e^{ba}/[{\rm const} + e^{ba}] \quad .$$
The downward deviations from the simple exponential increase of the Gompertz
law (1a) may become the smaller the better the data are (Gavrilova and Gavrilov 
2005).

Women should not be relied upon in tests of this exponential mortality 
increase. While men obey it quite nicely in Fig.1, the lawless women first 
have a mortality only about half as large as men, then increase their mortality
stronger than the men until finally the two mortalities become about equal 
near the age of 100 years. Thus if one plots female mortalities only above the 
age of 80 one sees strong curvature which can be misinterpreted as a mortality
deceleration away from the Gompertz law (1a). 

Germans also should not be relied upon. Wars may have long-time consequences, 
and German life expectancies in the 1960's decreased for a few years instead of 
the usual 
increase. The unification of East and West Germany in 1990 lead to womankind's
greatest birth strike in East Germany (now mostly over), with the number of 
children per women falling below one for some years. Sweden, in contrast,
avoided wars since two centuries, and Swedish life expectancy at birth also 
increased smoothly but non-linearly since two centuries, Fig.2, without showing
a reliable tendency to become constant. 

\begin{figure}[hbt]
\begin{center}
\includegraphics[angle=-90,scale=0.35]{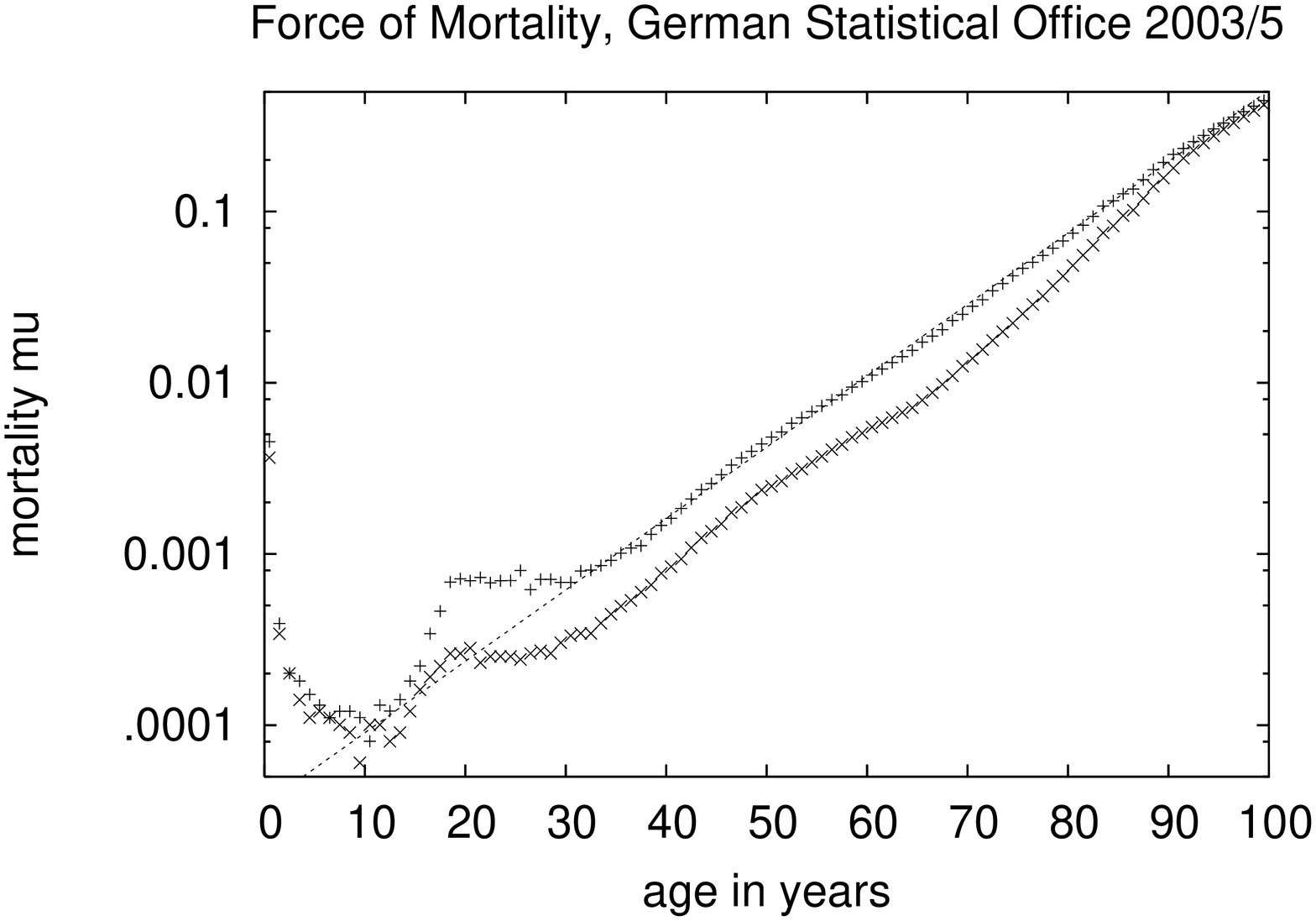}
\includegraphics[angle=-90,scale=0.35]{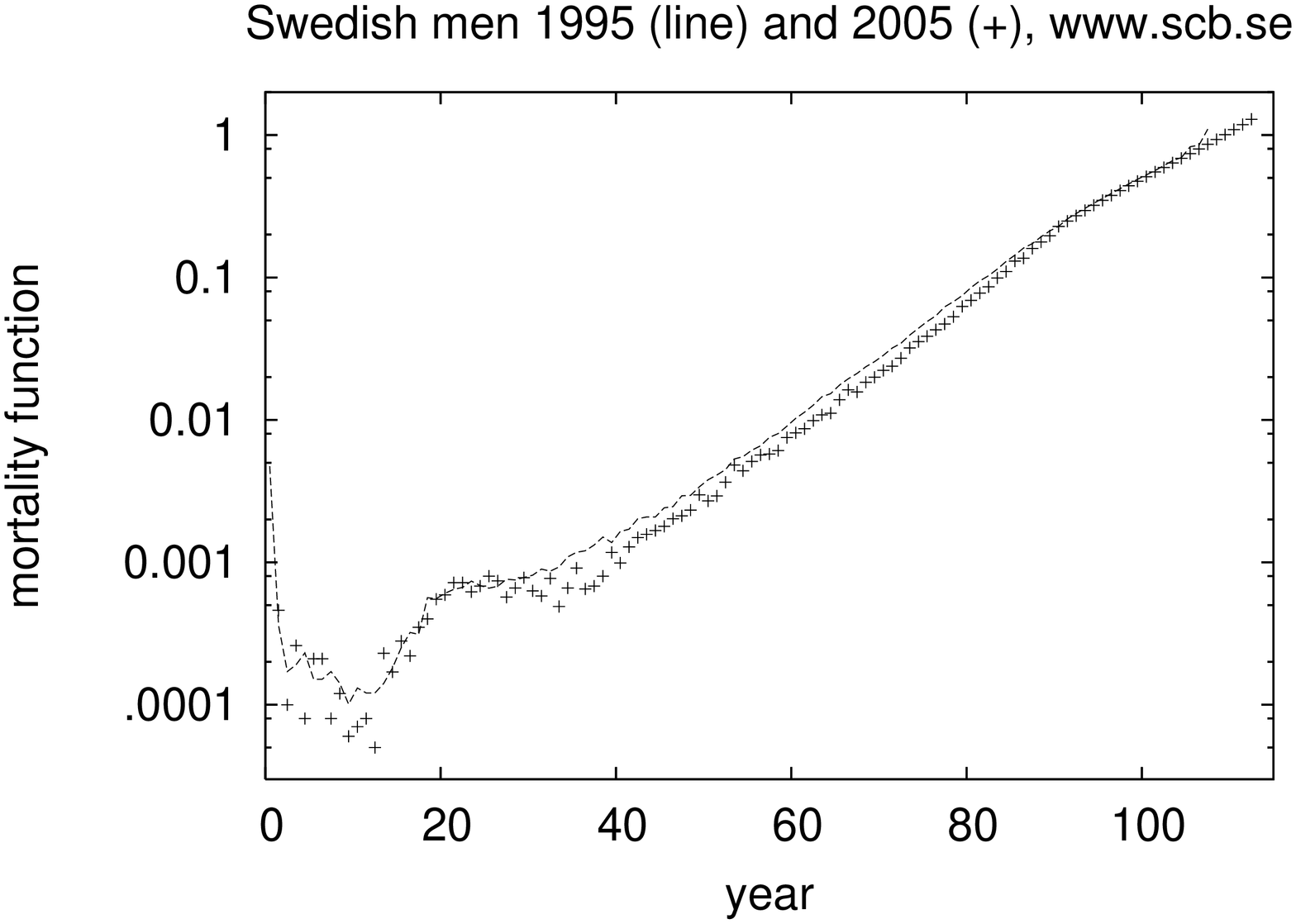}
\end{center}
\caption{Mortality function $\mu$ versus age for: a) German men (+) and women 
(x); b) Swedish men
}
\end{figure}
 
\begin{figure}[hbt]
\begin{center}
\includegraphics[angle=-90,scale=0.5]{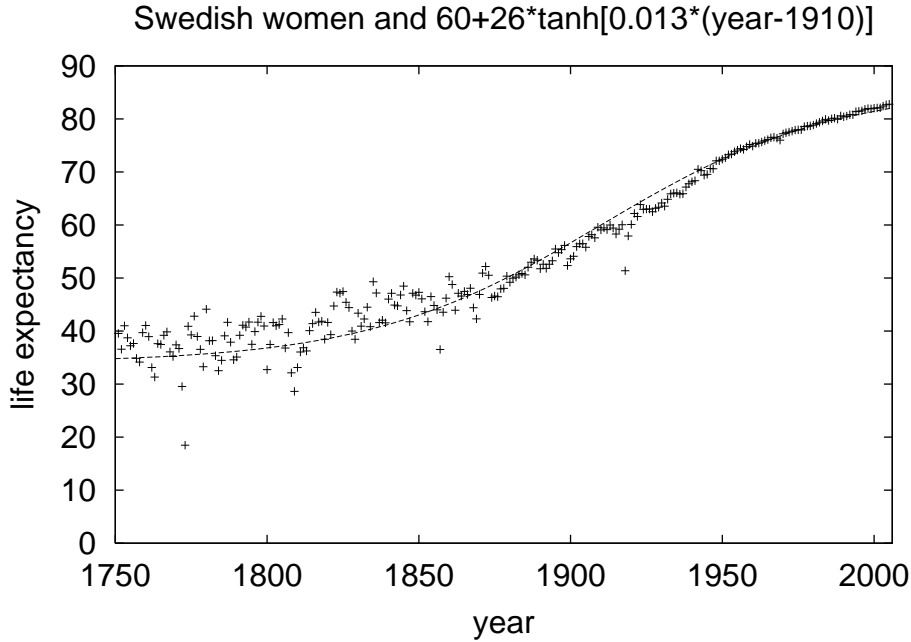}
\end{center}
\caption{Life expectancy at birth for Swedish women over 2.5 centuries, also
showing the improved statistical accuracy after the creation of a statistical 
central bureau in Sweden around 1860, The influenza pandemic in seen in 1918. 
}
\end{figure}
 
If one checks for each year in which country the life expectancy at birth is
highest, and then plots this life expectancy versus the calendar year, one sees 
roughly a straight line (Oeppen and Vaupel 2002) since 160 years, with Japan 
leading since many years (Cheung and Robine 2007). This linerarity was hardly 
valid around the year 1800, Fig. 2. 

If one compares different countries and different centuries, one sees some
universality, going back to Mildvan and Strehler but formulated in proper 
dimensionless form by Azbel 1996 (see also Gavrilov and Gavrilova 2001):
$$ \mu = {\rm const} \; b \; \exp[b(a-X)]   \eqno (2)$$
where $X \simeq 103$ years is a characteristic age (not a maximum age) for all 
human societies, independent of country and calendar years. Medical progress 
in the last two centuries increased the Gompertz slope $b$ but not the 
characteristic age $X$. However, during the last few decades one may have seen
a change to a constant $b$ and an increasing $X$ (Wilmoth et al 1999 and 2000, 
Yashin et al 2001, Edwards and Tuljapurkar 2005, Cheung and Robine 2007). 

Neither the latter changes not the downward deviations from the Gompertz law
are well established, and thus eq(2) is a good approximation for humans above
the age of 30 years.

At young age, drastic deviations are seen in Fig.1 from the straight line
symbolizing the Gompertz law (1a). Adding a constant to eq(1a) may
fit between 20 and 30 years, but still fails below 20 years. Mostly we will 
thus ignore child mortality and try to simulate models for adult mortality:
Don't trust anybody below 30.

In all our discussions, ``ageing'' is defined through the mortality, not 
through wisdom, beauty, health etc. Not only are mortalities much better 
documented and measurable than wisdom, they are also best suited for
computer simulations through population dynamics. A good aging model thus should
reproduce the empirical Gompertz law (1a), i.e. the exponential increase of 
adult mortalities with age. Refinements then should explain the mortality
minimum at childhood, and also the difference between male and female 
mortalities except for very old people. Once that is achieved one may apply 
the model for other questions like speciation or demography.

\section{Mutation Accumulation and the Penna Model}

More than half a century ago, Medawar tried to explain aging through the 
accumulation of inherited bad mutations over many generation. If such a 
hereditary disease kills a person at young age (before sexual maturity), that 
person has not produced any children, and this mutation will fail in Darwinian 
survival of the fittest. If, one the other hand, such a bad mutation kills this 
author at about the time this review is printed, then the government 
can save pension money. Thus bad mutations acting at young age can hardly 
spread in the population due to Darwinian selection pressure; for mutations
acting at old age this pressure is much weaker, and these mutations can spread
widely. This argument is not restricted to humans and indeed the Gompertz 
law also applies to many animals (Vaupel et al 1998), with different time scales
for the slope $b$ and the characteristic age $X$. 

Penna implemented this mutation accumulation hypothesis by dividing life into 
32 time intervals and by representing the genome (DNA) through a string of 32
bits, each of which can be zero or one. A zero bit means health, a bit set to 
one means a dangerous inherited disease starts to act from that age on which 
corresponds to the position of this bit in the bit-string. If $T$ (typically,
$T=3$) bits are active, their combined effect kills the individual. Each 
individual which has reached the minimum reproduction age of $R$ (typically, 
$R=8$) gets $B$ (typically, $1 \le B \le 4$) children at each time step,
where 32 time steps give the maximum life span. The child inherits the 
mother's genome except for $M$ (typically, $M = 1$) mutations of randomly 
selected bits where a zero bit becomes a one bit. (A bit set to one remains at
one in most versions of this model.) The number 32 is computationally 
convenient but biologically unrealistic; longer bit-strings are better and give 
roughly the same results if parameters are scaled properly ({\L}askiewicz et 
al 2005). In any case, parameters should be chosen such that in the stationary
equilibrium no individual reaches the age of 32 (or whatever the length of the 
bit-string is), since there it would die because of computational convenience
and not for biological reasons.

Starting with an ideal set of zero bits for all individuals, after about 
$10^3$ iterations the total population achieves a stationary value, and after
about $10^4$ iterations this is true also for the oldest individuals. Fig.3
shows large-scale simulations in reasonable agreement with the Gompertz law. 
Results do not strongly depend on the various parameters. However, it may happen
that for too low birth rates the whole population dies out (Malarz 2007);
this effect was also studied for a host-parasite system (Stauffer et al 2007).
For the oldest old the mortality may jump to infinity but this is hardly seen 
in simulations.

\begin{figure}[hbt]
\begin{center}
\includegraphics[angle=-90,scale=0.5]{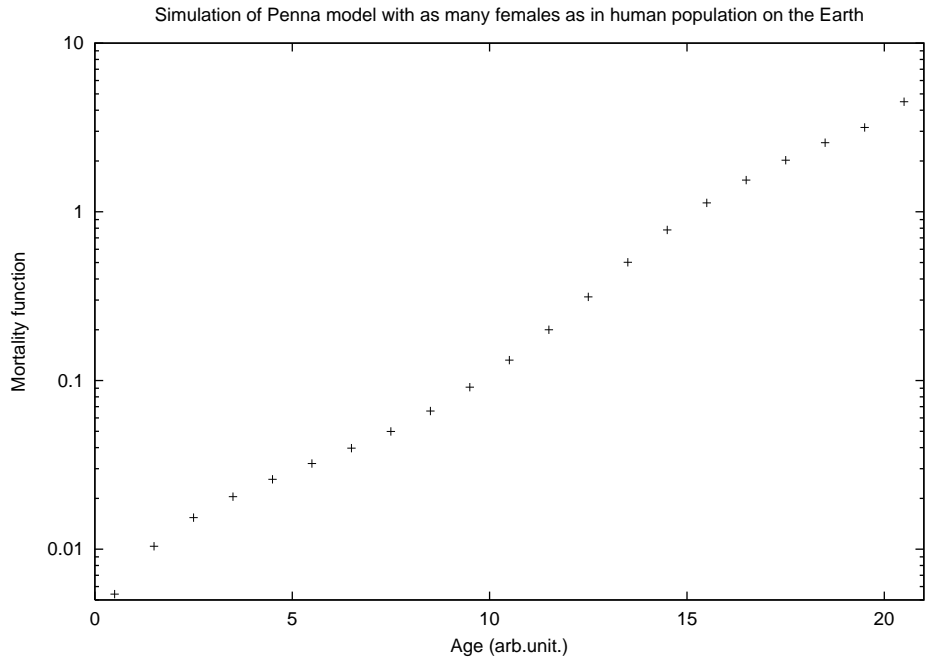}
\end{center}
\caption{Simulated mortality function for the asexual Penna model with 32 bits
and $R = 8$. From Stauffer et al (2006).
}
\end{figure}

The distribution of bits set to one shows a low probability for young age,
though $T-1$ randomly selected bit positions may have most bits set to one for
the whole population. Then, after the 
minimum reproduction age $R$, the fraction of mutated bits rises first slowly, 
then sharply, and reaches unity at some well-defined age which is about the 
maximum age in that population. For shorter times like $10^3$ iterations 
this maximum age is not yet sharply defined (Sitarz and Maksymowicz 2005).

\section{Applications of Asexual Model} 

To avoid a population growing exponentially to infinity, a Verhulst death
probability $N(t)/K$ is applied at each time step $t$, where $N$ is the 
current total population and $K$ is often called the carrying capacity, 
describing the limits of food and space. These deaths can be applied to 
everybody or (computationally more problematic but biologically more realistic)
to newborn babies only (Martins and Cebrat 2000). The latter choice is 
implemented indirectly if everybody is put on a lattice and offspring have to 
be placed on an empty lattice neighbor; then no Verhulst deaths are needed
at all.

Life means to eat and to be eaten. He et al (2005) put wolves, sheep and grass
onto a square lattice and let all animals age according to the Penna model. 
They found strong oscillations as is often the case in such prey-predator 
models, but also possible extinction of the wolves and even of the sheep. 

One biological observation regarded as crucial support for the mutation 
accumulation hypothesis are the Virginia opossum. These small mammals have
a low reproductive age and a low life expectation on the continent where
they are hunted by predators. But on islands without such predators, they 
get offspring later and die later. Such ``plasticity'' of the minimum age of 
reproduction is consistent with mutation accumulation but difficult to explain 
otherwise. recently, Reznick et al (2004) found the opposite effect for 
other animals. Fortunately, Altevolmer (1999) in between these biological 
observations simulated this plasticity in the Penna ageing model and found 
both decreases and increases of the minimum age of reproduction, depending 
on whether the predators kill mainly the young or mainly the old prey. Thus
simulations predicted some observations, and these observations do not 
contradict the mutation accumulation hypothesis in the Penna implementation.
 
If everybody in the present model has the same genome, then their genetic death
ages all agree, and the only variation would come from the Verhulst deaths (if 
these are not restricted to newborns). Biological experiments are sometimes 
made with inbred populations believed to have the same genes, but nevertheless
not all individuals die at the same age. This bad property of the present 
model was repaired by Pletcher and Neuhauser (2000) by combining it with
reliability theory (Gavrilov and Gavrilova 2001). Alternatively one may 
assume that too many active mutations do not kill automatically, but only 
with a probability depending on the difference of the active number of 
mutations and the threshold $T$ (Coe et al 2002).
 
\section{Sexual Penna Model}

Bacteria pretend not to have sex but sometimes they engage in exchange of 
genome. (``parasex''). More complicated organisms often divide into male and 
female individuals, and off-spring is created by combining parts of the 
paternal genome with parts of the maternal genome, using recombination. We
ignore some complications of reality at represent the paternal genome in a
sperm cell by one bit-string, and the maternal genome in the egg cell by
another bit-string. The child gets both bit-strings and thus has a diploid 
genome with two bit-strings of the same length. In reality these two bit-strings
correspond to two chromosomes, not to two nucleotide sequences on the DNA 
double helix (Bernardes 1996,  for a long review of early research see Moss de 
Oliveira et al. 1999). 

To create the paternal bit-string in the haploid sperm cell, the two paternal
bit-strings undergo with some probability (often taken as unity) a crossover 
(recombination): Some position on the bit-string is selected randomly, and all
the bits from on one side of this crossover point are taken from one of the two 
bit-string, and all the other bits from the other bit-string. In this way,
the sperm cell has a bit-string different from each of the two paternal 
bit-strings. An analogous random crossover creates the haploid genome of 
the egg cell. Then these two bit-strings fuse to form the diploid zygote from
which develops into the child. The resulting greater variety is supposed 
to justify the cost of sexual reproduction compared to bacterial cloning
but it is not clear that the profit outweighs this cost (Martins and 
Stauffer 2001): The city of Cologne later imposed a sex tax. Scharf (as cited
on page 91 of Stauffer et al 2006) offered the additional argument of 
pre-selection: Only the fastest sperm cell can fertilize the egg cell, and 
cells with bad mutations may move slower and thus are less likely to enter
the egg cell. This Darwinian preselection of the fittest sperm cell does not 
exist for asexual reproduction and may justify the existence of men. Fortunately
for me,  Mother Nature invented too late the trick to produce males without
mouth and stomach, suited only as fertilization machines with a high 
profit-to-cost ratio.

All men know that they die sooner than women because they are oppressed by them.
Scientific journals, however, feel obliged to publish also other reasons. One
blames Mother Nature: Women have two X chromosomes, and men have one X and 
one Y chromosome, with the Y chromosome containing only few genes. Using a 
sexual Penna model with many chromosomes, one of them the X (or Y), Schneider
et al (1998) found male mortalities to be about twice as high as female 
mortalities, except for the oldest old where the two roughly agree. 

A crucial test of this chromosome hypothesis would be life expectancies
for birds. While for mammals the males have different (XY) and the females the 
same (XX) chromosomes, for birds the situation is the opposite. Unfortunately 
two studies (Paevskii 1985; Austad 2001) contradict each other and no better 
ones are known to me. For humans the male-female differences in life 
expectancies are now usually several years (Oeppen and Vaupel 2002) and may be 
larger than the gains if cancer can be healed; thus understanding these
differences could be very helpful for human health. However, breeding, 
feeding, protecting and studying many thousand birds up to their intrinsic 
deaths seems not be to be as sexy as analyzing human genomes or claiming to
heal Alzheimer by stem cell research.

In any case, genetics alone can hardly explain {\it all} the differences 
in the life expectancies of men and women, since in Sweden this difference
changed rapidly over 250 years, Fig.4, too short a time to change appreciably
the human genome.

\begin{figure}[hbt]
\begin{center}
\includegraphics[angle=-90,scale=0.5]{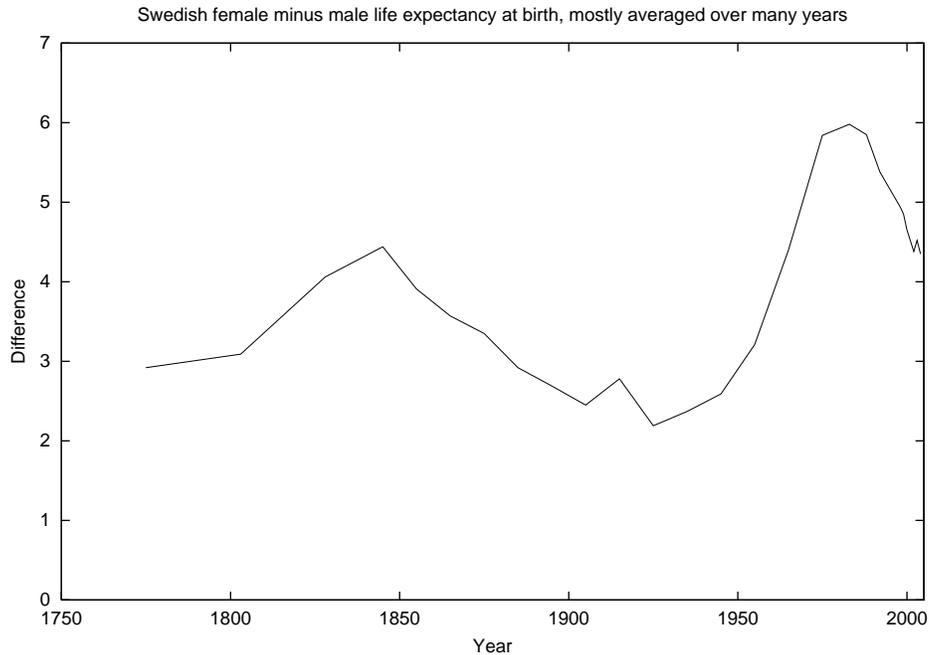}
\end{center}
\caption{Time variation of the advantages in life expectancies of women over 
men in Sweden. From Stauffer et al (2006) and www.scb.se.
}
\end{figure}

An important result of these sexual Penna model simulations is that menopause 
or its analogs, that means the cessation of female reproductive ability 
long before death, can emerge by itself as a result of the need for child care 
by the mother and an increase with age in the risk of giving birth. There is
no need of human culture and grandmothers teaching their grand children to
explain menopause, and indeed similar effects have been observed in many 
animals, not just in pilot whales as was thought long ago. Stauffer et al (2006)
discuss this point at length.
 
Sexual, as opposed to asexual, reproduction allows a new way to escape 
the detrimental effects of bad mutations. If we have two bit-strings of 
length 8 in the asexual case then 00010000 is clearly more favorable for 
survival than 11101111. However, for sexual reproduction most mutations are
recessive, i.e. the reduce the survival probability if and only if both bits 
are set at the same place in the two bit-strings. If these two bit-string 
examples form the diploid genome, and if all bit-positions are recessive, then 
the survival probability is not reduced at all, in spite if half the bits being
set to one. Thus it survival of the fittest may lead to individuals which have
half their bits set to one but with complimentary bit-strings in the two 
bit-strings of the haploid genomes.

This possibility was actually realized in models both from a genetics group 
(Zawierta et al 2007, Waga et al 2007) and from a physicist (P\c {e}kalski 
2007b) without aging. It also appears in the Penna model, if the recombination 
rate (crossover probability) is not equal to one as traditionally but is much 
smaller. We can measure this effect more quantitatively by the Hamming 
distance between the two bit-strings of the same individual. This Hamming 
distance counts how many bits are different in a position-by-position 
comparison of the two bit-strings. We compare only the bits up to the minimum
reproduction age $R$, since those at higher ages are mostly set to one anyhow.
Fig.5 shows for bit-strings of length 64 and $R = 40$ the normal behavior 
(x) at a rather high recombination probability $r = 0.128$; there is a 
single peak with a maximum at the rather low value of ten (of 40) bits: 
Evolution tried to reduce the number of mutations. At $r = 0$, on the other 
hand, the distribution splits up into two parts, one near the maximum of 40 
and one near zero. In most of the population one finds only
two different bit-strings which are complementary to each other. And most 
individuals have these two complimentary bit-strings in their diploid genome:
Evolution pushed for complimentary mutations, not for elimination of 
mutations. Waga et al (2007) used such concepts to simulate sympatric
speciation, but without the Penna ageing model.

\begin{figure}[hbt]
\begin{center}
\includegraphics[angle=-90,scale=0.5]{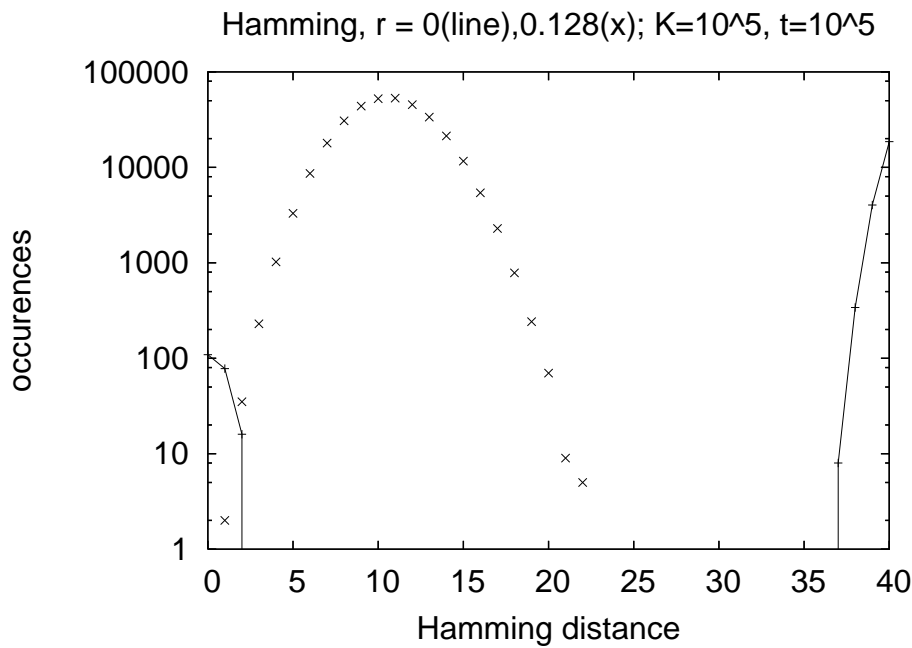}
\end{center}
\caption{Distribution of Hamming distances between the two bit-strings 
of each diploid genome with low (line) and intermediate (x) recombination
probabilities $r$. From Bo\'nkowska et al 2007.
}
\end{figure}
  
Such speciation means that one species slowly separates into two different
ones, without being separated by a geographical barrier. (Allopatric
speciation is caused by such a barrier, like homo sapiens Coloniensis and homo 
sapiens Neanderthalis being today separated by the Rhine river.) Luz-Burgoa
et al (2006) expanded the Penna ageing model by giving each individual 
besides the usual two-bit-string another pair of bit-strings, which are not 
related to ageing. Roughly, the number of bits set to one is called the 
phenotype. The Verhulst death probabilities depend on the phenotype, and
so does the mating preference: females prefer mating partners with the largest
(smallest) phenotype, if their own phenotype is larger(smaller) than average. 
Fig.1 of Luz-Burgoa et al shows how an initial single-peaked distribution of 
intermediate phenotypes divided after many iterations into two widely 
separated peaks at the largest and smallest possible phenotypes: Two species
emerged out of one, without any outside influence from two different 
food sources etc. (See P\c {e}kalski 2007a for a simplified model)

Practical applications include demography: How many young people have to 
support the upcoming retirement of this author? Bo\'nkowska et al (2006)
simulated this with a sexual Penna model, where each bit-string has a length
of 640 bits. Much simpler are the simulations (sections 6.1 and 9.5 in 
Stauffer et al 2006, Sumour et al 2007) where one age distribution of people
is iterated to give the age distribution in the next year, using realistic 
birth rates and Gompertz mortalities. Many European countries may need young
immigrants and increased retirement ages to deal with the reduced birth
rates and longer life expectancies of recent decades.

\section{Other Aging Simulations}

Weismann at the end of the 19th century suggested roughly that 
we die to make place for our children. In some sense, a simple model
(Stauffer and Radomski 2001) avoided all bit-string complications and had 
only two genetic properties for each individual: the minimum reproduction
age $R$ and the genetic death age $D$. Both mutated randomly up or down from 
one generation to the next. Without further restrictions the death age goes to 
infinity, nice but unrealistic. But if the birth rate was assumed to be roughly
inversely proportional to $D-R$, such that the expected number of offspring 
is constant and no longer influenced by mutations in $R$ and $D$, a stable
equilibrium was obtained: A finite death age is advantageous for the population
as a whole, in agreement with Weissmann. Modifications are needed to get a 
mortality increasing exponentially instead of linearly with age (Makowiec et 
al 2001).

Also quite simple are the models of Partridge and Barton, Jan, Dasgupta, 
Heumann and H\"otzel with two self-organizing mortalities, for juveniles and
adults. They finally lead to a rather realistic mortalities for many ages 
(Medeiros and Onody 2001). Later this model was used by Lobo and Onody (2006) 
to explain sex.

A reasonable mortality function was also obtained in a telomere model by
Masa et al (2006). The simplest and most recent model by Shklovskii (2005) does
not even require a computer simulation and uses reliability theory (Gavrilov
and Gavrilova 2001) for the immune system.

All these quantitative models, plus many earlier theories reviewed by Kirkwood
(2005) have not reached the popularity of the Penna model.

\bigskip
I thank all my collaborators in ageing research, in particular N. Jan, T.J.P. 
Penna, S. Moss de Oliveira, P.M.C. de Oliveira, A.T. Bernardes, S. Cebrat,
D. Makowiec, A. Maksymowicz, and A.O. Sousa (in temporal order). 

\section{References}
\parindent 0pt

Altevolmer AK. 1999. Virginia opossums, minimum reproduction age and predators 
in the Penna aging model. International Journal of Modern Physics 10: 717-721.

Austad SN. 2001. The Comparative Biology of Aging, in: Annual Review of
Gerontology and Geriatrics, Vol. 21, eds. V.J. Cristafalo and R. Adelman,
Springer, New York.

\medskip
Azbel MY. 1996. Unitary mortality law and species-specific age. Proceedings of 
the Royal Society of London B 263: 1449-1454.

\medskip
Bernardes AT. 1996. Strategies for reproduction and ageing. Ann. Physik 5: 
539-550.

\medskip
Bo\'nkowska K, Kula M, Cebrat S., Stauffer D, 2007. Inbreeding and outbreeding
depressions in the Penna model as a result of crossover frequency. International
Journal of Modern Physics 18 (8), in press.

\medskip
Cheung SLK, Robine JM. 2007. Increase in common longevity and the compression 
of mortality: The case of Japan. Population Studies 61: 85-97.

\medskip
Coe JB, Mao Y, Cates ME. 2002. Solvable senescence model showing a mortality 
plateau. Physical Review Letters 89: 288103. 

\medskip
Edwards RD, Tuljapurkar S. 2005. Inequality in life spans and a new perspective
on mortality convergence across industrialized countries. Population and
Development Review 31: 645-674, in particular Fig.6.

\medskip
Gavrilov LA, Gavrilova NS. 2001. The reliability theory of aging and longevity.
Journal of Theoretical Biology 213: 527-545.

\medskip
Gavrilova NS, Gavrilov LA 2005. Search for Predictors of Exceptional Human 
Longevity. In: ``Living to 100 and Beyond'' Monograph. The Society of 
Actuaries, Schaumburg, Illinois, USA, pp. 1-49.
http://library.soa.org/news-and-publications/publications/other-publications/monographs/pub-living- to-100-and-beyond-monograph.aspx

\medskip
He MF, Pan QH, Wang S. 2005. Final state of ecosystem containing grass, sheep 
and wolves with aging. International Journal of Modern Physics C 16: 177-190.

\medskip
Kirkwood TBL. 2005. Understanding the odd science of aging. Cell 120: 437-447.

\medskip
{\L}aszkiewicz A, Cebrat S,  Stauffer D. 2005. Scaling effects in the Penna 
ageing model. Advances in Complex Systems 8, 7-14.
  
\medskip
Lobo MP, Onody RN. 2006. Ploidy, sex and crossing over in an evolutionary 
aging model. Physica A 361: 239-249.

\medskip
Luz-Burgoa K. 2006. Thermodynamic behavior of a phase transition in a model for 
sympatric speciation.  Phys. Rev. E 74: 021910.

\medskip
Makowiec D, Stauffer D, Zielinski M. 2001. Gompertz law in simple computer 
model of aging of biological population. International Journal of Modern 
Physics C 12: 1067-1073.

\medskip
Malarz K. 2007. Risk of extinction - mutational meltdown or the
overpopulation, Theory in Bioscience 125: 247-156.

\medskip
Martins, J.S. S\'a, Cebrat S. 2000. Random deaths in a computational model for 
age-structured populations. Theory in Biosciences 119: 156-165.

\medskip
Martins, J.S. S\'a, Stauffer D. 2001. Justification of sexual reproduction by 
modified Penna model of ageing. Physica A 294, 191-194.

\medskip
Masa M, Cebrat S, Stauffer D. 2006. Does telomere elongation lead to a 
longer lifespan if cancer is considered? Physica A 364: 424-330.

\medskip
Medeiros NGF, Onody RN. 2001. Heumann-Hotzel model for aging revisited.
Physical Review E 64: 041915. 

\medskip
Moss de Oliveira S, de Oliveira PMC, Stauffer D. 1999. Evolution, Money,
War and Computers. Teubner, Leipzip and Stuttgart.

\medskip
Oeppen J, Vaupel JW. 2002. Demography - Broken limits to life expectancy.
Science 296: 1029-1031.

\medskip
Paevskii VA. 1985. Demography of Birds. Nauka, Moscow; in Russian.

\medskip
Partridge L. 2000. A singular view of ageing. Nature 447:262-263

\medskip
Penna TJP. 1995. A bit-string model for biological aging. Journal of 
Statistical Physics 78: 1629-1633 

\medskip
P\c {e}kalski A. 2007a. Role of Selective Mating in Population Splitting.
International Journal of Modern Physics C 18 (9), in press.

\medskip
P\c {e}kalski A. 2007b. Simple Model of Mating Preference and Extinction Risk. 
International Journal of Modern Physics C 18 (10), in press.

\medskip
Reznick DN, Bryant MJ, Roff D, Ghalambor CK, Ghalambor DE. 2005. Effect of 
extrinsic mortality on the evolution of senescence in guppies.  Nature 431: 
1095-1099.

\medskip
Robine JM, Vaupel JW. 2001. Supercentenarians: slower ageing individuals or 
senile elderly? Experimental Gerontology  36: 915-930.

\medskip
Schneider J, Cebrat S, Stauffer D. 1998. Why do women live longer than men?
International Journal of Modern Physics C 9: 721-725.

\medskip
Shklovskii BI. 2005. A simple derivation of the Gompertz law for human 
mortality. Theory in Biosciences 123: 431-433. 

\medskip
Sitarz M, Maksymowicz A. 2005. Divergent evolution paths of different genetic 
families in the Penna model. International Journal of Modern Physics C 16: 
1917-1925.

\medskip
Stauffer D, Moss de Oliveira S, de Oliveira PMC, S\'a Martins JS. 2006.
Biology, Sociology, Geology by Computational Physicists. Amsterdam, Elsevier.

\medskip
Stauffer D, Proykova A, Lampe K. 2007. Monte Carlo Simulation of Age-Dependent 
Host-Parasite Relations, Physica A, in press, doi:10.1016/j.physa .2007.05.028.

\medskip
Sumour MA, El-Astal AH, Shabat MM, Radwan MA. 2007. Simulation of Demographic 
Change in Palestinian Territories. International Journal of Modern Physics C 18 
(1?), in press.

\medskip
Thatcher AR, Kannisto V, Vaupel JW. 1998. The Force of Mortality
at Ages 80 to 120. Odense University Press, Odense.

\medskip
Vaupel JW et al. 1998. Biodemographic trajectories of longevity. Science 280:
855-860.

\medskip
Waga, W, Mackiewicz, D, Zawierta M, Cebrat S. 2007, Sympatric speciation as
intrinsic property of the expanding population.  Theory in Biosciences,
in press; doi:10.1007/s12064-007-0010-z. 

\medskip
Wilmoth JR, Deegan LJ, Lundstr\"om H, Horiuchi S: Increase of maximum life-span 
in Sweden 1861-1999. 2000. Science 289: 2366-2368; Wilmoth JR, Horiuchi S. 1999.
Rectangularization revisited: Variability of age at death within human 
populations. Demography 36: 475-490.

\medskip
Yashin AI, Begun AS, Boiko SI, Ukraintseva SV, Oeppen J. 2001. The new trends 
in survival improvement require a revision of traditional gerontological 
concepts. Experimental Gerontology 37: 157-167.

\medskip
Zawierta M, Biecek P, Waga W, Cebrat S. 2007. The role of
intragenomic recombination rate in the evolution of population's genetic pool.
Theory in Biosciences 125: 123-132 (2007).
\end{document}